\documentclass[12pt]{article}
\usepackage[a4paper, top=16mm, text={170mm, 248mm}, includehead, includefoot, hmarginratio=1:1, heightrounded]{geometry}
\usepackage{amsmath,amssymb,mathrsfs,amsthm,tikz,shuffle,paralist}
\usepackage{dsfont}
\usepackage{color}
\definecolor{darkred}{RGB}{173,34,48}

\usepackage[all]{xypic}
\usepackage{jheppub}
\usepackage{hyperref}

\title{A Note on Letters of Yangian Invariants}
\author[a,b,c,d]{Song He}
\author[a,d]{Zhenjie Li}
\affiliation[a]{CAS Key Laboratory of Theoretical Physics, Institute of Theoretical Physics, Chinese Academy of Sciences, Beijing 100190, China}
\affiliation[b]{
School of Fundamental Physics and Mathematical Sciences, Hangzhou Institute for Advanced Study, UCAS, Hangzhou 310024, China}
\affiliation[c]{ICTP-AP
International Centre for Theoretical Physics Asia-Pacific, Beijing/Hangzhou, China}
\affiliation[d]{School of Physical Sciences, University of Chinese Academy of Sciences, No.19A Yuquan Road, Beijing 100049, China}
\emailAdd{songhe@itp.ac.cn}
\emailAdd{lizhenjie@itp.ac.cn}

\date{\today}

\abstract{
Motivated by reformulating Yangian invariants in planar ${\cal N}=4$ SYM directly as $d\log$ forms on momentum-twistor space, we propose a purely algebraic problem of determining the arguments of the $d\log$'s, which we call ``letters", for any Yangian invariant. These are functions of momentum twistors $Z$'s, given by the positive coordinates $\alpha$'s of parametrizations of the matrix $C(\alpha)$, evaluated on the support of polynomial equations $C(\alpha) \cdot Z=0$. We provide evidence that the letters of Yangian invariants are related to the cluster algebra of Grassmannian $G(4,n)$, which is relevant for the symbol alphabet of $n$-point scattering amplitudes. For $n=6,7$, the collection of letters for all Yangian invariants contains the cluster ${\cal A}$ coordinates of $G(4,n)$. We determine algebraic letters of Yangian invariant associated with any ``four-mass" box, which for $n=8$ reproduce the $18$ multiplicative-independent, algebraic symbol letters discovered recently for two-loop amplitudes. 
}

\begin{document}

\maketitle

\section{Introduction}

Recent years have witnessed tremendous progress in unravelling hidden mathematical structures of scattering amplitudes, especially in planar ${\cal N} = 4$ supersymmetric Yang-Mills theory (SYM) ({\it c.f.} \cite{
ArkaniHamed:2012nw,Arkani-Hamed:2013jha}). Moreover, formidable progress has been made in computing multi-loop scattering amplitudes of the theory, most notably by the hexagon and heptagon bootstrap program~\cite{Dixon:2011pw, Dixon:2014xca, Dixon:2014iba, Drummond:2014ffa, Dixon:2015iva, Caron-Huot:2016owq}. A crucial assumption for the bootstrap is that the collection of the letters entering the {\it symbol} (which is the maximally iterated coproduct of generalized polylogarithms), or the {\it alphabet}, consists of only $9$ and $42$ cluster variables of $G(4,n)$~\cite{Golden:2013xva}. These are dual conformally invariant (DCI), rational functions of momentum twistors describing the kinematics of $n$-point amplitudes~\cite{Hodges:2009hk}. 
The space of generalized polylogarithms with correct alphabet is remarkably small, and certain information from physical limits and symmetries suffices to fix the result (see~\cite{Caron-Huot:2020bkp} for a recent review). The symbol of the six-point amplitude (or hexagon) has been determined through seven and six loops for MHV and NMHV cases respectively~\cite{Caron-Huot:2019vjl}, and similarly that of the seven-point amplitude (or heptagon) has been determined through four loops for these cases respectively~\cite{Dixon:2016nkn, Drummond:2018caf}. 

Starting $n=8$, the cluster bootstrap becomes intractable since the Grassmannian cluster algebra~\cite{Scott2006} becomes infinite type. It is an important open question how to find a finite subset of cluster variables which appear in the symbol alphabet of multi-loop amplitudes, and there has been significant progress {\it e.g.} by studying their Landau singularities \cite{Dennen:2015bet, Dennen:2016mdk, Prlina:2017azl,Prlina:2018ukf}. Moreover, it is known that starting $n=8$, {\it algebraic} (irrational) functions of momentum twistors appear in the alphabet, as one can already see in four-mass box integrals needed for one-loop amplitudes with $k\geq 2$ (such algebraic letters are also present for higher-loop MHV amplitudes with $n\geq 8$). 

Recently, a new computation for two-loop $n=8$ NMHV amplitude is performed using the so-called ${\bar Q}$ equations~\cite{CaronHuot:2011kk}, which provides a candidate of the symbol alphabet for $n=8$; in particular, it provides the space of $18$ algebraic letters which involve square roots of the kinematics. 
After that new proposals concerning the symbol alphabet for $n\geq 8$ have appeared~\cite{Drummond:2019cxm, Henke:2019hve,Arkani-Hamed:2019rds}, which exploit mathematical structures such as tropical Grassmannian~\cite{speyer2005tropical} and positive configuration space~\cite{Arkani-Hamed:2020cig}. 

In this short note, we make a simple observation that an alternative construction, based on solving polynomial equations associated with plabic graphs for Yangian invariants, may provide another route to symbol alphabet including algebraic letters. We will argue that the {\it alphabet} of Yangian invariants, which will be defined shortly, contains not only the (rational) symbol letters of $n=6$ and $n=7$ amplitudes but also the algebraic letters for two-loop $n=8$ NMHV amplitude mentioned above. 

We emphasize that generally for any Yangian invariant, the alphabet includes but is not restricted to the collection of poles, which have been explored in the literature so far, {\it e.g.} in the context of the so-called cluster adjacency \cite{Drummond:2017ssj,Drummond:2018dfd,Drummond:2018caf,Mago:2019waa,Golden:2019kks,Gurdogan:2020tip}. As we will see already for $n=7$,  by going through all Yangian invariants we find $42$ poles in total, which are $7$ frozen variables and $35$ unfrozen ones, {\it i.e.} $7$ of the $42$ cluster ${\cal A}$ variables of $G(4,7)$ are still missing! On the other hand, the alphabet of all possible Yangian invariants consists of $63$ letters, which include the $42$ cluster variables. The search for alphabets of Yangian invariants is by itself a beautiful open problem related to positive Grassmannian, plabic graphs and cluster algebra~\cite{ArkaniHamed:2012nw}, and let's first formulate the general problem as suggested by~\cite{pricom}. 

An N${}^k$MHV Yangian invariant is an on-shell function of (super-)momentum twistors $\{Z |\eta\}_{1\leq a\leq n}$, which is associated with a certain $4k$-dimensional positroid cell of the positive Grassmannian $G_+(k,n)$~\footnote{One can easily extend our construction to general $m$ but in this note we only consider $m=4$.}; the cell can be parametrized by a $k\times n$  matrix $C(\{\alpha\}_{1\leq i\leq 4k})$ where $\alpha_i$'s are positive coordinates, and the on-shell function (which has Grassmann-degree $4k$) is given by
\begin{equation}
    Y (\{Z | \eta\})=\int \prod_{i=1}^{4k} d\log \alpha_i~\delta^{4k |4k } (C(\{\alpha\}) \cdot (Z | \eta) )\,,
\end{equation}
where the contour of the integral encloses solutions of equations $\sum_{a=1}^n C_{I, a} (\{\alpha\})  Z_a=0$ (for $I=1,2,\cdots, k$). As first proposed in~\cite{Arkani-Hamed:2017vfh} and studied further in~\cite{He:2018okq}, it is instructive to write Yangian invariants, or more general on-shell (super-)functions, as {\it differential forms} on momentum-twistor space, where we replace the fermionic variables $\eta_a$ by the differential $d Z_a$ for $a=1,\cdots, n$. Remarkably, after the replacement the Yangian invariant defined as a $4k$ form, ${\cal Y} (\{Z | dZ\}):=Y|_{\eta_a \to dZ_a}$, becomes the pushforward of the canonical form of the cell, $\prod_i d\log \alpha_i$, to $Z$ space:
\begin{equation}
{\cal Y}^*(Z| dZ)=\{\bigwedge_{i=1}^{4k} d \log \alpha^*_i (Z) \}, \quad {\it with}~\{\alpha^* (Z)\}~{\it solutions~of}~C(\alpha)\cdot Z=0\,.
\end{equation}
In other words, we evaluate the form on solutions, $\alpha^*(Z)$, of the polynomial equations $C(\alpha)\cdot Z=0$, and we have used subscript $*$ as a reminder that there can be multiple solutions in which case ${\cal Y}^*$ becomes a collection of $4k$ forms, one for each solution (the number of solutions, $\Gamma(C)$ is given by the number of isolated points in the intersection $C^{\perp} \cap Z$~\cite{ArkaniHamed:2012nw}).  The solutions, $\alpha^*(Z)$, are GL$(1)$-invariant functions of Pl\"{u}cker coordinates on $G(4,n)$, $\langle a\,b\,c\,d\rangle:={\rm det} (Z_a Z_b Z_c Z_d)$, and when there are multiple solutions, $\Gamma(C)>1$, they are generally algebraic functions.

Since symbol letters of loop amplitudes are related to arguments and indices of the involved generalized polylogarithms, it is only natural to refer to the arguments of the $d\log$'s on the support of $C \cdot Z=0$, as the {\it letters} of a given Yangian invariant. One can define the {\it alphabet} of a Yangian invariant to be the collection of possible letters, {\it i.e.} the solutions $\{\alpha_i^*(Z)\}$, which form a $4k \times \Gamma(C)$ matrix. However, there are ambiguities since under a reparametrization of the cell, $C(\{\alpha\}) \to C(\{\alpha'\})$, which leaves the form invariant, the alphabet generally changes. Such reparametrizations are related to cluster transformations acting on variables of (positroid cells of) $G_+(k,n)$~\cite{ArkaniHamed:2012nw}. In principle one could attempt to find all possible letters for a given positroid cell, by scanning through all transformations that leave the canonical form invariant, but the resulting collection of letters can easily be infinite for $n\geq 8$ ! 

Therefore, we restrict ourselves to a very small subset of such transformations, namely those generated by equivalence moves acting on {\it plabic graphs} associated with a positroid cell. Therefore, we define the {\it alphabet} of any given Yangian invariant to be the collection of face or edge variables of all possible plabic graphs.  It is obvious that this collection of letters is finite for any Yangian invariant, and we can further take the union of the alphabets for all Yangian invariants of the same $n$ and $k$, which will be referred to as the alphabet of a given $n$ and $k$. 

There is still a residual redundancy in this definition: any monomial reparametrization of the form $\alpha'_j=\prod_i \alpha_i^{n_{i,j}}$ for $1\leq j\leq d$ with $\det n_{i,j}=1$ leaves the form invariant. Even for a given plabic graph, we have such redundancy in the definition of face or edge variables, and the invariant content of the alphabet is the space generated by monomials of $\{\alpha^*(Z)_i\}$ (or the linear space spanned by $\{\log \alpha^*_i\}$),  and one can choose some representative letters which give the same space. For example, when $\alpha'(Z)$ are rational with $\Gamma(C)=1$, one can always choose a set of irreducible polynomials of Pl\"{u}cker coordinates, which are analogous to ${\cal A}$ coordinates of cluster algebra of $G(4,n)$. Similar choices can be made for algebraic solutions with $\Gamma(C)>1$, modulo ambiguities from the fact that monomials of these algebraic letters can form rational letters which are irreducible polynomials. 

In our definition, letters of Yangian invariants are generally not DCI, thus they are analogous to the cluster ${\cal A}$- coordinates. It is trivial to homogenize them and obtain the analog of cluster ${\cal X}$ coordinates. For our main example of algebraic letters, where the Yangian invariants correspond to leading singularities of four-mass boxes, a natural way to define ${\cal X}$-like variables is by considering ratio of conjugates involving the square roots. We find that the collection of such ratios turns out to give precisely the same $18$ algebraic alphabet for two-loop NMHV amplitudes in \cite{Zhang:2019vnm}. 

\section{Letters of Yangian invariants}

In the following, we will illustrate our algorithm by finding (subsets of) alphabets of Yangian invariants for certain $n$ and $k$. The procedure goes as follows: for any given $n$ and $k$, we first scan through all possible $4k$-dimensional cell of $G_+(k,n)$ and list the resulting Yangian invariants. The representation of any Yangian invariant can be obtained using the procedure given in~\cite{ArkaniHamed:2012nw}, in terms of the matrix $C(\{\alpha\})$ with {\it canonical coordinates} $\{\alpha\}_{1\leq i\leq 4k}$ for the cell; equivalently these coordinates can be identified with non-trivial edge variables of a representation plabic graph. Then one can start to apply square moves to the graph, which generate reparametrization of the cell and possibly find more letters. In simple cases, it is straightforward to find all equivalence moves and obtain the complete alphabet of the Yangian invariant; for more involved cases, we will content with ourselves in finding a subset of the alphabet by applying such move once, and the results turn out to be already illuminating. 

\subsection{Letters of NMHV and $\overline{\rm MHV}$ invariants}
First, we present the simplest non-trivial Yangian invariant, which is the only type for NMHV ($k=1$). A $4$-dimensional cell $C\in G_+(1,n)$ can be parametrized as
\[
	(\dots,0,1,0\dots,0,\alpha_1,0,\dots,0,\alpha_2,0,\dots,0,\alpha_3,0,
	\dots,0,\dots,0,\alpha_4,0,\dots),
\]
where only $a$, $b$, $c$, $d$ and $e$-th entry are non-zero, and the on-shell function is given by the R invariant $[a,b,c,d,e]$. The solution of the linear equation $C \cdot Z=0$ is 
simply 
\begin{equation}
	\alpha_1=-\frac{\langle acde\rangle}{\langle bcde\rangle},\,\,
	\alpha_2=-\frac{\langle abde\rangle}{\langle cbde\rangle},\,\,
	\alpha_3=-\frac{\langle abce\rangle}{\langle dbce\rangle},\,\,
	\alpha_4=-\frac{\langle abcd\rangle}{\langle ebcd\rangle},
\end{equation}
and one can easily check that ${\cal Y}_{n,1}(a,b,c,d,e)=\prod_{i=1}^4 d\log \alpha_i^*(Z)$ is indeed the form obtained by replacing $\eta$ by $dZ$ in $[a,b,c,d,e]$. Obviously any reparametrization is trivial in this case, and we can choose representative letters to be the five Pl\"{u}cker coordinates appeared above. 

By taking the union of alphabets of all NMHV invariants for $n$ points,  we trivially get the collection of all Pl\"{u}cker coordinates. Also note that in this special case the representative letters are exactly the poles of the Yangian invariant, but we will see immediately that this is no longer true beyond NMHV. 

Our next example is for $\overline{\rm MHV}$ ($k=n{-}4$), where the unique Yangian invariant for $n$ points is given by the top cell of $G_+(n{-}4,n)$ (or equivalently that of $G_+(4,n)$, and the form is the familiar cyclic measure:
$$
{\cal Y}_{n, n-4}=\frac{d^{4 n} Z/({\rm vol~GL}(4) )}{(1234)(2345) \cdots (n123)}\,,
$$
The poles of the invariant ($\overline{\rm MHV}$ amplitude) are the frozen Pl\"{u}cker coordinates, as they should be. However, the letters for any plabic graph involve variables other than these frozen ones. Let's start with $n=6$ (for $n=5$ it is just NMHV invariant $[1,2,3,4,5]$), and a representative plabic graph is
\begin{center}
    \includegraphics[scale=0.5]{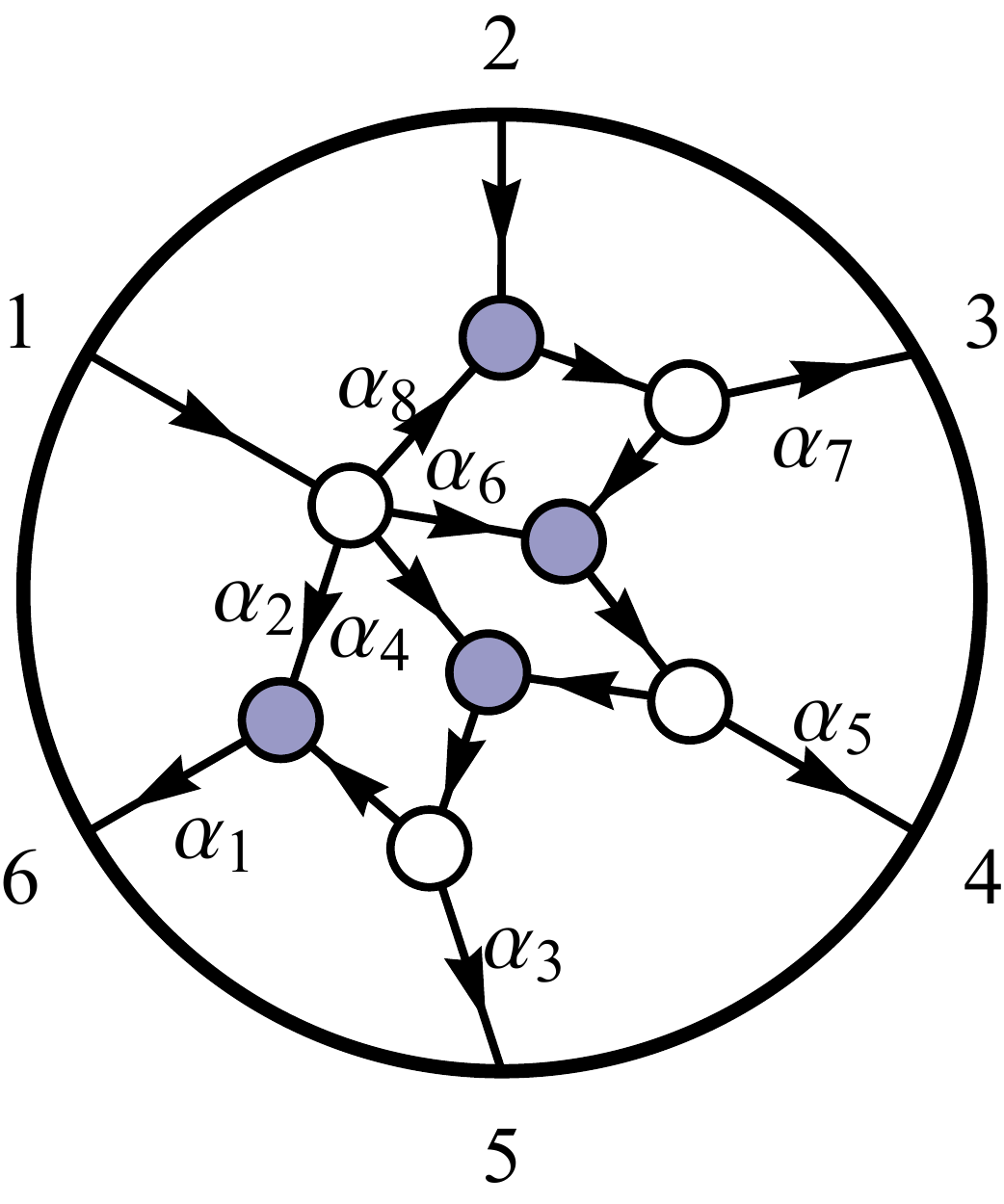}
\end{center}
The corresponding $C$ matrix reads 
\[
\begin{pmatrix}
 1 & \alpha _2+\alpha _4+\alpha _6+\alpha _8 & \left(\alpha _2+\alpha _4+\alpha _6\right) \alpha _7 & \left(\alpha _2+\alpha _4\right) \alpha _5 & \alpha _2 \alpha _3 & 0 \\
 0 & 1 & \alpha _7 & \alpha _5 & \alpha _3 & \alpha _1 
\end{pmatrix}
\]
and the solution of $C(\alpha) \cdot Z=0$ is
\begin{align*}
    &\alpha_1= \frac{\langle 2345\rangle}{\langle 3456\rangle},
    \alpha_2= -\frac{\langle 1234\rangle \langle 3456\rangle}{\langle 2345\rangle \langle 2346\rangle},
    \alpha_3= -\frac{\langle 2346\rangle}{\langle 3456\rangle},
    \alpha_4= -\frac{\langle 1236\rangle \langle 3456\rangle}{\langle 2346\rangle \langle 2356\rangle},\\
    &\alpha_5= \frac{\langle 2356\rangle}{\langle 3456\rangle},
    \alpha_6= -\frac{\langle 1256\rangle \langle 3456\rangle}{\langle 2356\rangle \langle 2456\rangle},
    \alpha_7= -\frac{\langle 2456\rangle}{\langle 3456\rangle},
    \alpha_8= -\frac{\langle 1456\rangle}{\langle 2456\rangle},
\end{align*}
where we find the letters given by the following $9$ Pl\"{u}cker coordinates:
\begin{equation}
\langle 1234 \rangle, 
\langle 2346 \rangle, 
\langle 2345 \rangle, 
\langle 1236 \rangle, 
\langle 2356 \rangle, 
\langle 1256 \rangle, 
\langle 2456 \rangle, 
\langle 1456 \rangle, 
\langle 3456 \rangle.
\end{equation}
Note that these are the ${\cal A}$ coordinates of a cluster in $G(4,6)$: $6$ frozen variables plus $3$ unfrozen ones, and they can be arranged as the quiver below. This quiver can be obtained as the dual of the plabic graph~\cite{ArkaniHamed:2012nw}, and in this case, any square moves on plabic graphs are in 1:1 correspondence with mutations on the cluster. Thus, without explicitly performing square moves, we see the complete alphabet of the Yangian is given by the collection of all ${\cal A}$ coordinates, which are the $15$ Pl\"{u}cker coordinates. 
\[
\xymatrix{
    \fbox{$\langle 2345\rangle$}\ar[rd]&&&&\\
    &\langle 2346\rangle\ar[d]\ar[r]&\langle 2356\rangle\ar[d]\ar[r]&\langle 2456\rangle\ar[d]\ar[r]&\fbox{$\langle 3456\rangle$}\ar[d]\\
    &\fbox{$\langle 1234\rangle$}&\fbox{$\langle 1236\rangle$}\ar[lu]&\fbox{$\langle 1256\rangle$}\ar[lu]&\fbox{$\langle 1456\rangle$}\ar[lu]\\
}
\]

For general $n$, one can always choose a representative plabic graph, which corresponds to an initial cluster of $G(4,n)$ (as a generalization of the quiver above). It is straightforward to see that for such a graph, the letters are given by $4(n-4)+1$ Pl\"{u}cker coordinates, with $n$ frozen variables and $3(n-5)$ unfrozen ones. The square moves still correspond to mutations but only those special ones acting on a node with two incoming and two outgoing arrows (dual to a square face), and such mutations generate new letters that are still Pl\"{u}cker coordinates.

\subsection{Letters of $n=6$ and $n=7$ invariants}

From the discussions above, one can determine the collection of letters for all possible Yangian invariants for $n=6$. In fact, either from the union of alphabets for NMHV invariants or from that of N${}^2$MHV one, we get $9+6=15$ ${\cal A}$ coordinates of $G(4,6)$.

Now we move to $n=7$, where in addition to NMHV Yangian invariants, we also have two classes of N${}^2$MHV ones, which are the first cases with letters other than Pl\"{u}cker coordinates. Since N${}^2$MHV invariants are given by parity-conjugate of the NMHV ones for $n=7$, the poles of the former can be obtained as the parity conjugate of poles of the latter (which are Pl\"{u}cker coordinates).  It is straightforward to see that parity conjugate of $\langle i i{+}1 j j{+}1\rangle$ and $\langle i{-}1 i i{+}1 j \rangle$ are proportional to Pl\"{u}cker coordinates of the same type, thus for $n=7$ the only kind of Pl\"{u}cker coordinates with non-trivial parity conjugate are $\langle 1 3 4 6\rangle$ and cyclic permutations; under parity, we have $\langle 7(12)(34)(56)\rangle$ {\it etc.}, and we see that $35$ unfrozen cluster variables of $G(4,7)$ appear as poles of Yangian invariants. However, we will see new letters appearing for N${}^2$MHV invariants, which are not parity conjugate of Pl\"{u}cker coordinates, and they are crucial for getting all cluster variables in this case. 

A representative plabic graph of a Yangian invariant in the first class is
\begin{center}
    \includegraphics[scale=0.5]{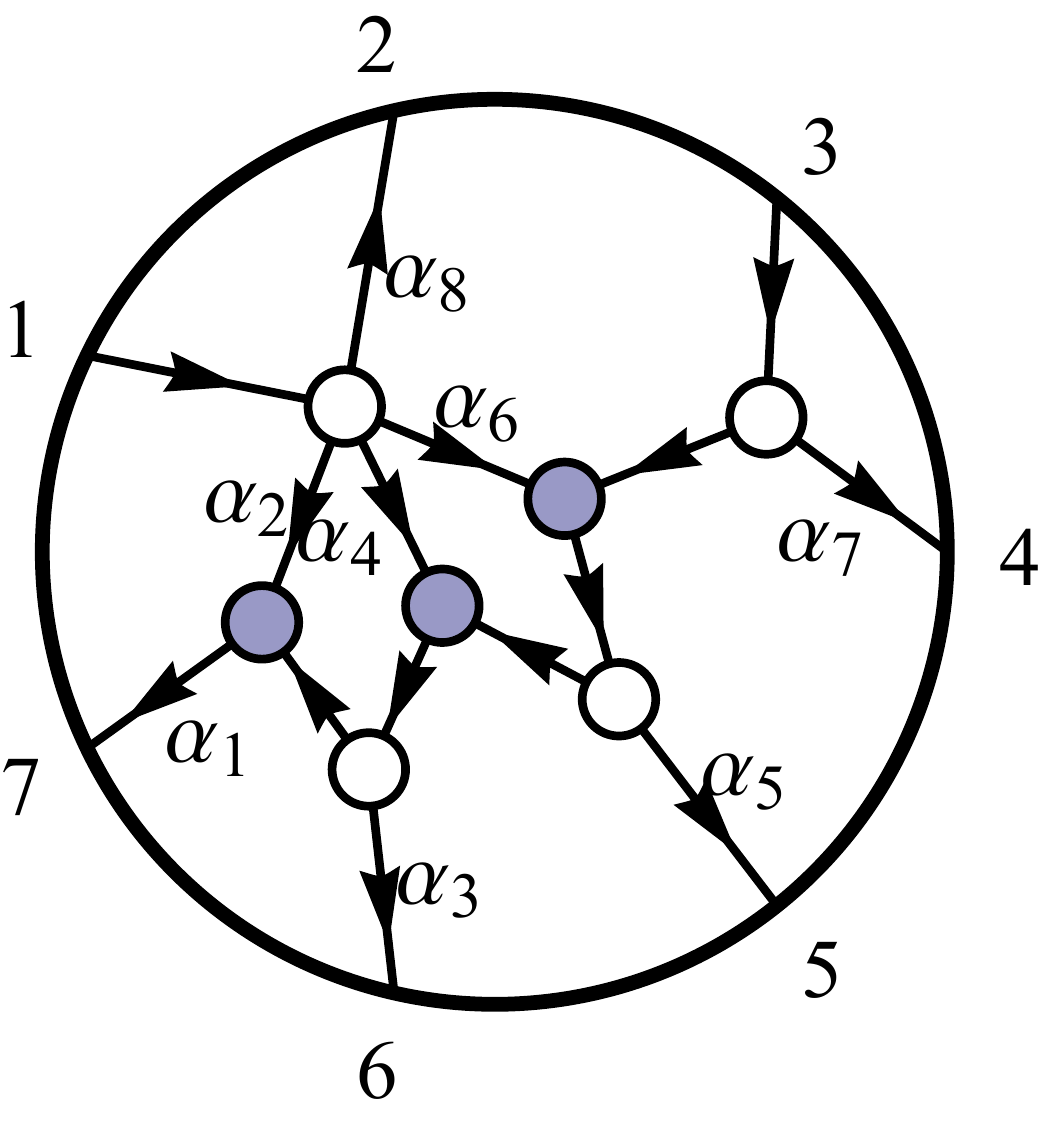}
\end{center}    
The corresponding $C$ matrix reads
\[
\begin{pmatrix}
 1 & \alpha _8 & \alpha _2+\alpha _4+\alpha _6 & \left(\alpha _2+\alpha _4+\alpha _6\right) \alpha _7 & \left(\alpha _2+\alpha _4\right) \alpha _5 & \alpha _2 \alpha _3 & 0 \\
 0 & 0 & 1 & \alpha _7 & \alpha _5 & \alpha _3 & \alpha _1
\end{pmatrix}
\]
and the solution of $C(\alpha) \cdot Z=0$ is 
\begin{align*}
    &\alpha_1= \frac{\langle 3456\rangle}{\langle 4567\rangle},
    \alpha_2= -\frac{\langle 4567\rangle \langle 5(12)(34)(67)\rangle}{\langle 2567\rangle \langle 3456\rangle \langle 3457\rangle},
    \alpha_3= -\frac{\langle 3457\rangle}{\langle 4567\rangle},
    \alpha_4= -\frac{\langle 4567\rangle \langle 7(12)(34)(56)\rangle}{\langle 2567\rangle \langle 3457\rangle \langle 3467\rangle},\\
    &\alpha_5= \frac{\langle 3467\rangle}{\langle 4567\rangle},
    \alpha_6= -\frac{\langle 1267\rangle \langle 4567\rangle}{\langle 2567\rangle \langle 3467\rangle},
    \alpha_7= -\frac{\langle 3567\rangle}{\langle 4567\rangle},
    \alpha_8= -\frac{\langle 1567\rangle}{\langle 2567\rangle},
\end{align*}
where we have defined
\[
    \langle a(b c)(d e)(f g)\rangle \equiv\langle a b d e\rangle\langle a c f g\rangle-\langle a b f g\rangle\langle a c d e\rangle\,.
\]
Again since we are interested in polynomials of Pl\"{u}cker coordinates, the letters for this plabic graph are the following $10$:
\begin{equation}
\begin{aligned}
&\langle 1267 \rangle, 
\langle 1567 \rangle, 
\langle 2567 \rangle, 
\langle 3456 \rangle,
\langle 3457 \rangle,
\langle 3467 \rangle,\\ 
&\langle 3567 \rangle, 
\langle 4567 \rangle, 
\langle 5 (67)(34)(12)\rangle,
\langle 7 (56)(34)(12)\rangle.
\end{aligned}
\end{equation}
By considering cyclic rotations of the labels, we see $7$ new letters of the form $\langle i\,(i \,i{+}1) (i{+}2\,i{+}3) (i{-}2\,i{-}1)\rangle$ for $i=1, \cdots, 7$, in addition to the $7+28$ Pl\"{u}cker coordinates. However,  if we compare to the $42$ (unfrozen) ${\cal A}$ coordinates of $G(4,7)$, $7$ variables of the form $\langle i (i{-}1\,i{+}1) (i{+}2\,i{+}3) (i{-}3\,i{-}2)\rangle$ (for $i=1,\cdots, 7$) are still missing. Can we find such cluster variables as letters of this Yangian invariant?

Very nicely, we will see that by applying square moves, these new letters appear in equivalent plabic graphs for the same Yangian invariant. Recall that it is more convenient to describe the square moves using face variables, which are given by monomials of our coordinates (certain edge variables)~\cite{ArkaniHamed:2012nw}. For an internal face with face variable $f$, after a square move, the only new factor introduced in the new set of face or edge variables is given by $1+f$. 

We consider square move on either of the two internal faces (the left one is adjacent to $\alpha_2$, $\alpha_4$, and the right one is adjacent to $\alpha_4$, $\alpha_6$); their face variables are
\[
    f_{1}=\frac{\alpha_4}{\alpha_2}=\frac{\langle 3456\rangle\langle 7(12)(34)(56)\rangle}{ \langle 3467\rangle\langle 5(12)(34)(67)\rangle},\quad
    f_{2}=\frac{\alpha_6}{\alpha_4}=\frac{\langle 1267\rangle \langle 3457\rangle}{\langle 7(12)(34)(56)\rangle}.
\]
After performing the square move, we obtain the following new factors,
\begin{equation}
    1+f_{1}=-\frac{\langle 3457\rangle \langle 6(12)(34)(57)\rangle}{\langle 3467\rangle \langle 5(12)(34)(67)\rangle},\quad 
    1+f_{2}=\frac{\langle 1257\rangle \langle 3467\rangle}{\langle 7(12)(34)(56)\rangle},
\end{equation}
and we see that a new letter $\langle 6(12)(34)(57)\rangle$ appears. This letter and its cyclic permutations exactly give the missing $7$ cluster variables mentioned above!

A representative plabic graph of a Yangian invariant in the second class is
\begin{center}
    \includegraphics[scale=0.5]{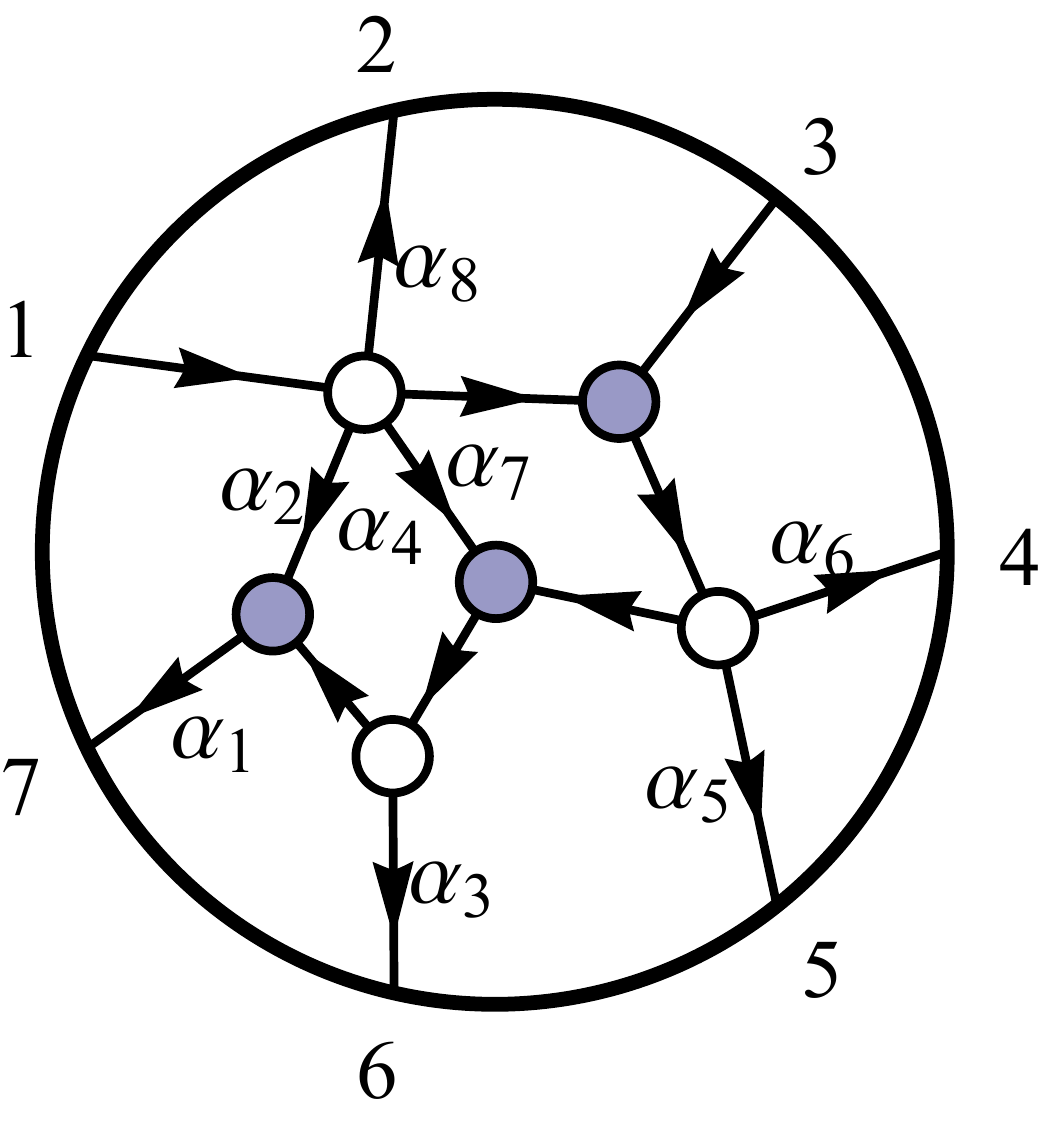}
\end{center}
The corresponding $C$ matrix reads
\[
\begin{pmatrix}
 1 & \alpha _8 & \alpha _2+\alpha _4+\alpha _7 & \left(\alpha _2+\alpha _4\right) \alpha _6 & \left(\alpha _2+\alpha _4\right) \alpha _5 & \alpha _2 \alpha _3 & 0 \\
 0 & 0 & 1 & \alpha _6 & \alpha _5 & \alpha _3 & \alpha _1
\end{pmatrix}
\]
and the solution of $C(\alpha) \cdot Z=0$ is 
\begin{align*}
    &\alpha_1= \frac{\langle 3456\rangle}{\langle 4567\rangle},
    \alpha_2= -\frac{\langle 4567\rangle \langle 3(12) (45) (67)\rangle}{\langle 2367\rangle \langle 3456\rangle \langle 3457\rangle},
    \alpha_3= -\frac{\langle 3457\rangle}{\langle 4567\rangle},\\
    &\alpha_4= -\frac{\langle 1237\rangle \langle 4567\rangle}{\langle 2367\rangle \langle 3457\rangle},
    \alpha_5= \frac{\langle 3467\rangle}{\langle 4567\rangle},
    \alpha_6= -\frac{\langle 3567\rangle}{\langle 4567\rangle},
    \alpha_7= \frac{\langle 1267\rangle}{\langle 2367\rangle},
    \alpha_8= -\frac{\langle 1367\rangle}{\langle 2367\rangle}.
\end{align*}
We obtain the following letters, which also miss the last class of $7$ cluster variables:
\begin{equation}
\langle 1237 \rangle, 
\langle 1267 \rangle, 
\langle 1367 \rangle, 
\langle 2367 \rangle,
\langle 3456 \rangle, 
\langle 3457 \rangle, 
\langle 3467 \rangle, 
\langle 3567 \rangle, 
\langle 4567 \rangle, 
\langle 3 (12)(45)(67)\rangle.
\end{equation}
Similarly we can apply square moves of either of the following two internal faces (the left is adjacent to $\alpha_2$, $\alpha_4$, and the right one is adjacent to $\alpha_4$, $\alpha_7$) with variables 
\[
    f_{1}=\frac{\alpha_4}{\alpha_2}=\frac{\langle 1237\rangle\langle 3456\rangle}{\langle 3(12)(45)(67)\rangle},\quad
    f_{2}=\frac{\alpha_7}{\alpha_4}=-\frac{\langle 1267\rangle \langle 3457\rangle}{\langle 1237\rangle \langle 4567\rangle},
\]
and the new factors obtained from such moves are given by
\begin{equation}
    1+f_{1}=\frac{\langle 1236\rangle\langle 3457\rangle}{\langle 3(12)(45)(67)\rangle},\quad 
    1+f_{2}=-\frac{\langle 7(12)(36)(45)\rangle}{\langle 1237\rangle \langle 4567\rangle}.
\end{equation}
Thus we find yet another new letter $\langle 7(12)(36)(45)\rangle$, and together with cyclic permutations, they give $7$ new variables which are not cluster variables!  In fact, by considering all possible plabic graphs for these Yangian invariants, we obtain an alphabet that consists of all $42$ unfrozen cluster variables of $G(4,7)$ (plus $7$ frozen ones), as well as $14$ new letters that are not cluster variables (all from the second type of invariants); $7$ in the cyclic class of $\langle 7(12)(36)(45)\rangle$ and $7$ in the class of $\langle 7(14)(23)(56) \rangle$. These $63$ letters make up the complete alphabet for $n=7$ Yangian invariants. 

\subsection{Algebraic letters of N${}^2$MHV invariants}

Finally, we present our main example of algebraic letters, namely leading singularities of four-mass boxes (see the left figure below). Without loss of generality, we consider that of the four-mass box $(a,b,c,d)=(1,3,5,7)$ for $n=8$ (the right figure below; the other four-mass box $(a,b,c,d)=(2,4,6,8)$ is given by cyclic rotation by $1$).

\begin{center}
    \begin{tikzpicture}[scale=1.8]
        \node[fill=black,circle,draw=black, inner sep=0pt,minimum size=10pt] at (-2,0) {};
        \node[fill=black,circle,draw=black, inner sep=0pt,minimum size=10pt] at (-2,-1) {};
        \node[fill=black,circle,draw=black, inner sep=0pt,minimum size=10pt] at (-1,-1) {};
        \node[fill=black,circle,draw=black, inner sep=0pt,minimum size=10pt] at (-1,0) {};
        \draw[line width=0.4mm] (-2,0) -- (-2,-1) -- (-1,-1) -- (-1,0) -- cycle;\draw[line width=0.4mm] (-2.6,-0.8) -- (-2,-1);\draw[line width=0.4mm] (-2,-1) -- (-1.8,-1.6);\draw[line width=0.4mm] (-1,-1) -- (-0.4,-0.8);\draw[line width=0.4mm] (-1,-1) -- (-1.2,-1.6);\draw[line width=0.4mm] (-2.6,-0.2) -- (-2,0);\draw[line width=0.4mm] (-1.8,0.6) -- (-2,0);\draw[line width=0.4mm] (-1.2,0.6) -- (-1,0);\draw[line width=0.4mm] (-1,0) -- (-0.4,-0.2);
        \node at (-1.2,0.8) {$a$};
        \node at (0,-0.2) {$b{-}1$};
        \node at (-0.2,-0.8) {$b$};
        \node at (-1.2,-1.8) {$c{-}1$};
        \node at (-1.8,-1.8) {$c$};
        \node at (-3,-0.8) {$d{-}1$};
        \node at (-2.8,-0.2) {$d$};
        \node at (-1.8,0.8) {$a{-}1$};
        \node at (-2.4,0.1) {$\cdot$};\node at (-2.3,0.3) {$\cdot$};\node at (-2.1,0.4) {$\cdot$};\node at (-0.9,-1.4) {$\cdot$};\node at (-0.7,-1.3) {$\cdot$};\node at (-0.6,-1.1) {$\cdot$};\node at (-0.6,0.1) {$\cdot$};\node at (-0.7,0.3) {$\cdot$};\node at (-0.9,0.4) {$\cdot$};\node at (-2.4,-1.1) {$\cdot$};\node at (-2.3,-1.3) {$\cdot$};\node at (-2.1,-1.4) {$\cdot$};
    \end{tikzpicture}
    \includegraphics[scale=0.5]{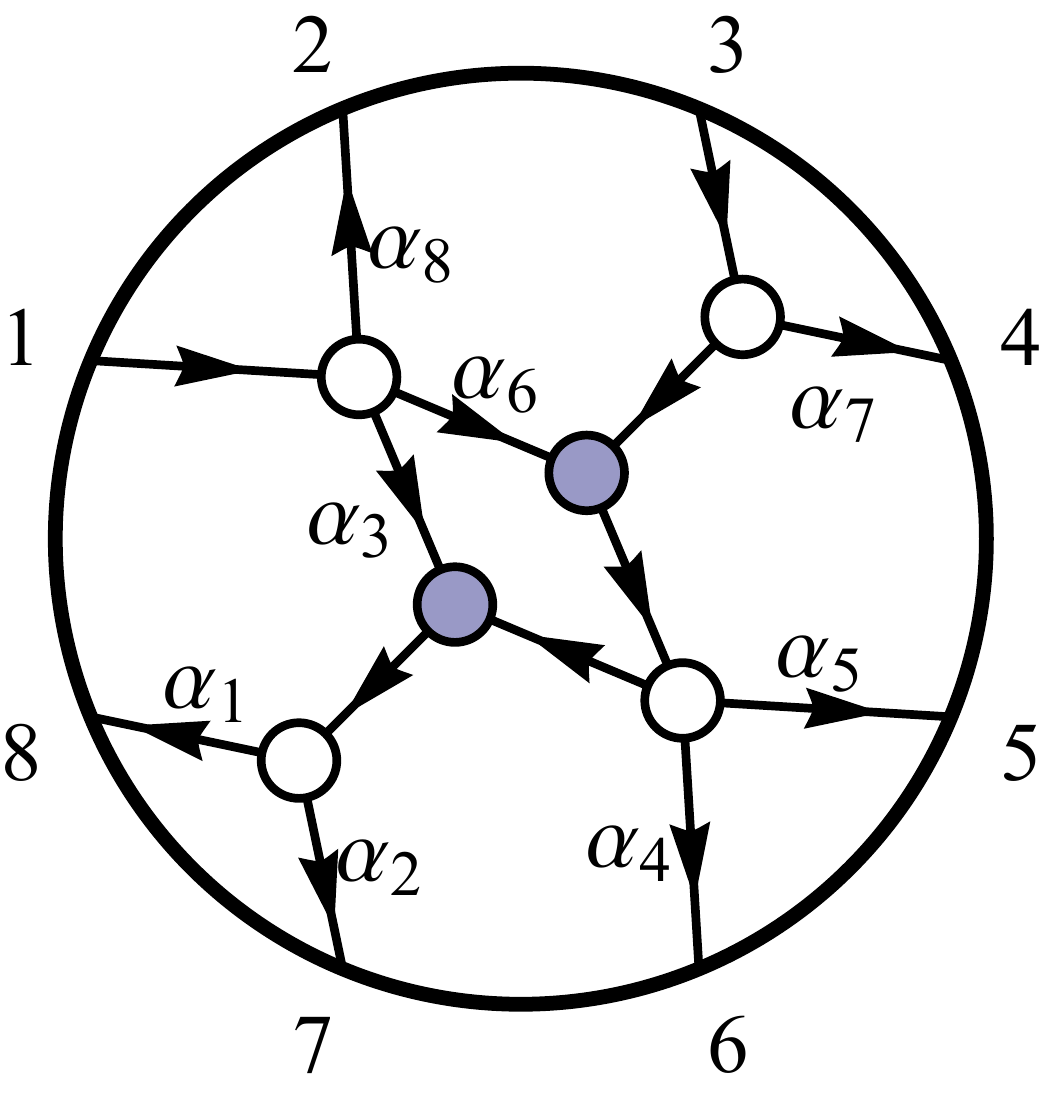}
\end{center}
For this representative plabic graph, the corresponding $C$ matrix reads
\[
\begin{pmatrix}
 1 & \alpha _8 & \alpha _3+\alpha _6 & \left(\alpha _3+\alpha _6\right) \alpha _7 & \alpha _3 \alpha _5 & \alpha _3 \alpha _4 & 0 & 0 \\
 0 & 0 & 1 & \alpha _7 & \alpha _5 & \alpha _4 & \alpha _2 & \alpha _1
\end{pmatrix}
\]
and the solution of $C(\alpha)\cdot Z=0$ is
\begin{align*}
    &\alpha_1 = -\frac{\langle 3456\rangle \langle 127B\rangle}{\langle 456A\rangle \langle 128B\rangle},
    \alpha_2 = \frac{\langle 3456\rangle}{\langle 456A\rangle},
    \alpha_3 = -\frac{\langle 456A\rangle \langle 128B\rangle}{\langle 3456\rangle \langle 278B\rangle},
    \alpha_4 = -\frac{\langle 345A\rangle}{\langle 456A\rangle},\\
    &\alpha_5 = \frac{\langle 346A\rangle}{\langle 456A\rangle},
    \alpha_6 = \frac{\langle 1278\rangle}{\langle 278B\rangle},
    \alpha_7 = -\frac{\langle 356A\rangle}{\langle 456A\rangle},
    \alpha_8 = -\frac{\langle 178B\rangle}{\langle 278B\rangle}
\end{align*}
where the two twistors $A, B$ are parametrized as
\[
    A= Z_7+\frac{\alpha_1}{\alpha_2} Z_8 =: Z_7+\alpha Z_8,\quad
    B=Z_3+\alpha_7 Z_4 =:Z_3+\beta Z_4,
\]
then $\alpha$, $\beta$ satisfy $\langle 12AB\rangle=0$ and $\langle 56AB\rangle=0$, {\it i.e.}
\begin{equation}
    \alpha=-\frac{\langle 5673\rangle+\langle 5674\rangle \beta}{\langle 5683\rangle+\langle 5684\rangle \beta},\quad 
    \beta=-\frac{\langle 1237\rangle+\langle 1238\rangle \alpha}{\langle 1247\rangle+\langle 1248\rangle \alpha},
\end{equation}
and we have two solutions for these quadratic equations, which we denote as  $X_+$ and $X_-$ for $X=A,B$. This is the source of letters involving square roots, and the discriminant of the quadratic equations, $\Delta$, reads
\[
\Delta:=(\langle (34) \cap (127) 568 \rangle +\langle (34) \cap (128) 567 \rangle)^2 - 4 \langle 
 7 (12) (34) (56) \rangle \langle 8 (12) (34) (56) \rangle\,.
\]
All the algebraic letters for this invariant involve the square root of $\Delta$. Therefore, we have the following $2$ Pl\"{u}cker coordinates and $8$ ${\cal A}$-like algebraic letters: 
\begin{equation}\label{algletters}
\langle 3456\rangle, \langle 1278\rangle, \langle 456 A\rangle, \langle 356 A\rangle, \langle 346 A\rangle, \langle 345A \rangle, \langle 128 B\rangle, \langle 127 B\rangle, \langle 178 B\rangle, \langle 278 B\rangle\,.
\end{equation}
For each algebraic letter, one can plug in two solutions, $X_+$ and $X_-$ but they are not multiplicative independent since the product is a rational function (which is expected to be given by rational letters from other Yangian invariants), thus we have $8$ independent algebraic letters for this plabic graph. 

Quite nicely, by a square move on the internal face, we obtain yet another algebraic letter. The face variable of the internal face is
\begin{equation}
    f=\frac{\alpha_6}{\alpha_3}=-\frac{\langle 1278\rangle\langle 3456\rangle}{\langle 456A\rangle \langle 128B\rangle},
\end{equation}
and the corresponding square move produces a new factor in the numerator:
\begin{equation}\label{lastletter}
    1+f=\frac{\langle 456A\rangle \langle 128B\rangle-\langle 1278\rangle\langle 3456\rangle}{\langle 456A\rangle \langle 128B\rangle}=\frac{\langle (AB)\cap (456) 128\rangle}{\langle 456A\rangle \langle 128B\rangle}\,.
\end{equation}
Therefore, from the graph and the one after square move, we find $9$ algebraic letters: the $8$ in \eqref{algletters} and $\langle (AB)\cap (456) 128\rangle$ in \eqref{lastletter}, which are all ${\cal A}$-like variables. It is straightforward to check that these $9$ algebraic letters generate the same space as the $9$ algebraic letters with this $\Delta$~\cite{Zhang:2019vnm}, modulo some rational letters. In fact, these $9$ letters provide a basis for the algebraic alphabet found in~\cite{Zhang:2019vnm}.

In fact, we can find all possible algebraic letters for this Yangian invariant by cyclic symmetry. The four-mass box is invariant under cyclic rotation by $2$, and one can apply it to the $8$ algebraic letters in \eqref{algletters}, which take the form $\langle A i j k\rangle$ and $\langle B i' j' k'\rangle$ for $i, j, k\in \{3,4,5,6\}$ and $i', j', k' \in \{7,8,1,2\}$. Under the rotation, we obtain the following $8$ letters: $\langle A'\,i+2\,j+2\,k+2\rangle$, $\langle B'\,i'+2\,j'+2\,k'+2\rangle$ 
with
\[
    A'=Z_1+\alpha' Z_2,\quad B'=Z_5+\beta' Z_6,
\]
and $\alpha'$ and $\beta'$ are generated by cyclic rotation by $2$ of $\alpha$ and $\beta$ respectively (they share the same square root). Thus, even without performing square moves explicitly, we find $16$ distinct algebraic letters (if we rotate by $2$ again, no new letters appear); one can check that $9$ out of the $16$ letters are multiplicative independent.

Alternatively, we can write the algebraic letters similar to cluster ${\cal X}$ variables, which are dual conformally invariant. In fact, in the above example, we already see two DCI letters which are internal face variable $f$ and the one after square move, $1+f$. Recall the well-known variables $z$ and $\bar{z}$ defined by the equations:
 \begin{equation}\label{zzbar}
z\bar{z}=\frac{\langle 1278\rangle\langle3456\rangle}{\langle1256\rangle\langle 3478\rangle}\:,\:
        (1-z)(1-\bar{z})=\frac{\langle 1234\rangle\langle5678\rangle}{\langle1256\rangle\langle 3478\rangle}\,.
        \end{equation}
It is easy to see that $f=z/(1-z)$ and $1+f=1/(1-z)$. 

To find ${\cal X}$-like letters more systematically, let's consider DCI ratios of the form $f(X_+)/f(X_-)$, where $f(X)$ is any of the $16$ ${\cal A}$-like letters with $X=A,B,A',B'$. This is a natural way to find $16$ ${\cal X}$-like letters, which are not multiplicative independent and the rank is $9$ as expected. Together with the other half from the box $(2,4,6,8)$, these ${\cal X}$-like letters generate precisely the same space of DCI algebraic letters given in~\cite{Zhang:2019vnm}. These new representation of the same space of algebraic letters are good for studying multiplicative relations, which have been systematically derived and summarized for any multiplicity in~\cite{He:2020vob}.
 
Since this is the only type of $n$-point N$^2$MHV algebraic Yangian invariants, we have exhausted algebraic letters for $k=2$: for each four-mass box $(a,b,c,d)$, there are $9$ independent algebraic letters which share the same square root.

\section{Discussions}

In this note, we have proposed an algebraic problem of finding the collection of letters, or arguments of $d\log$'s, of any Yangian invariant using parameterizations given by plabic graphs. For a given plabic graph, the $4 k$ polynomial equations $C(\alpha) \cdot Z=0$ provide a map from $4 k$-dimensional cell of $G_+(k,n)$, parametrized by $4k$ $\alpha$'s, to a collection of functions defined on $G(4,n)$; the alphabet of a Yangian invariant consists of such functions for all plabic graphs related to each other by square moves. In addition to alphabet of NMHV and $\overline{\rm MHV}$ invariants which are just Pl\"{u}cker coordinates, we show that for $n=6,7$, the union of alphabets for all invariants includes all cluster ${\cal A}$ variables of G$(4,n)$, and for $n=8$ the algebraic letters of N${}^2$MHV invariants coincide with the $18$ multiplicative-independent symbol letters found for two-loop NMHV amplitudes~\cite{Zhang:2019vnm}.

We have only provided some data for the general problem, and it would be highly desirable to have a general method for constructing letters for any plabic graph, without the need of case-by-case study. This may involve some algorithm directly in $Z$ space, which mimics the construction of cells in $G_+(k,n)$ from the map $C\cdot Z=0$~\cite{ArkaniHamed:2012nw}. Such a construction would circumvent the bottleneck of our computation, namely the need to scan through all plabic graphs of a Yangian invariant. 

Relatedly, another pressing question here is how to systematically understand the transformations of letters under square moves of the plabic graphs. As we have seen in the top-cell ($\overline{\rm MHV}$) case, these transformations correspond to certain cluster transformations, and we would like to understand the analog of these for lower-dimensional cells. The appearance of non-cluster variables in the alphabet of certain Yangian invariant is interesting. By studying all possible types of (rational) Yangian invariants with $k=2$~\cite{ArkaniHamed:2012nw}, we find that only two of them, the one we have seen above for $n=7$ and one for $n=8$, contain such non-cluster variables. It would be interesting to understand the origin of these non-cluster variables. 

Going into fine structures of the alphabet, one can ask ``cluster-adjacency" kind of questions not for the poles of a Yangian invariant, but rather the letters which appear in the $d\log$ form. That is, what letters can appear as arguments of a $4k$-dim $d\log$ form? When the letters are all cluster variables, we expect them to be in the same cluster, but we will need some new ideas when algebraic letters (and those non-cluster variables) are involved. The answer to this kind of questions may provide new insights for the study of symbol alphabet of multi-loop amplitudes. 

Last but not least, we would like to study this problem at higher $n$ and $k$. Already for $n=8$, it would be interesting to find the complete alphabet show that the union of alphabets include all the $180$ rational letters of~\cite{Zhang:2019vnm}. For $n=9$, we find that the $9\times 9$ independent algebraic letters from the $9$ N${}^2$MHV four-mass boxes are not enough for just the two-loop NMHV amplitudes~\cite{He:2020vob}. This is not surprising since there are two types of algebraic Yangian invariants for $k=3$ (both with $\Gamma(C)=2$), for which we have not been able to determine the alphabet yet. We leave all these fascinating open questions to future investigations. \\

{\bf Notes added}: During the preparation of the manuscript, \cite{Mago:2020kmp} appeared on arXiv which has some overlap with the result presented in this note. 

\section*{Acknowledgements} We are grateful to N. Arkani-Hamed for motivating us to study this problem and for inspiring discussions. We thank Chi Zhang for collaborations on related projects. All the plabic graphs in this paper are made using the Mathematica package \texttt{positroids.m} of J. Bourjaily~\cite{Bourjaily:2012gy}. This work is supported in part by Research Program of Frontier Sciences of CAS under Grant No. QYZDBSSW-SYS014 and National Natural Science Foundation of China under Grant No. 11935013.\\
\bibliographystyle{utphys}
\bibliography{yangian}

\end{document}